\documentstyle[amsfonts,preprint,aps]{revtex}

\input epsf

\renewcommand{\vec}[1]{\mbox{\boldmath$#1$}}

\begin{document}

\draft

\title{Pauli Blocking of Collisions in a Quantum Degenerate Atomic Fermi Gas}

\author{B. DeMarco, S. B. Papp, and D. S. Jin \cite{adr1}}
\address{JILA, \\ National Institute of Standards and Technology
and University of Colorado
\\ and \\ Physics Department, University of Colorado, Boulder, CO
80309-0440} \date{\today} \maketitle

\begin{abstract}

We have produced an interacting quantum degenerate Fermi gas of atoms
composed of two spin-states of magnetically trapped $^{40}$K. The
relative Fermi energies are adjusted by controlling the population in
each spin-state. Measurements of the thermodynamics reveal the
resulting imbalance in the mean energy per particle between the two
species, which is as large as a factor of 1.4 at our lowest
temperature. This imbalance of energy comes from a suppression of
collisions between atoms in the gas due to the Pauli exclusion
principle. Through measurements of the thermal relaxation rate we
have directly observed this Pauli blocking as a factor of two
reduction in the effective collision cross-section in the quantum
degenerate regime.

\end{abstract}


The experimental realization of a quantum degenerate Fermi gas of
atoms \cite{sciarticle} introduces a novel tool for the study of
quantum phenomena. The initial work, which extended the evaporative
cooling and magnetic trapping techniques used for atomic
Bose-Einstein condensation, produced a nearly ideal Fermi gas.
Although the lack of interparticle interactions makes theoretical
interpretation straightforward and may facilitate precision
measurements \cite{fermiCs}, a richer spectrum of phenomena can arise
in a dilute, interacting system.  Predicted phenomena for an
interacting atomic Fermi gas include zero sound \cite{zero},
suppression of elastic and inelastic collisions
\cite{ourpra,suppress}, component separation \cite{phase}, and the
prospect of a paired state at low temperature\cite{phase,BCS},
similar to Bardeen-Cooper-Schrieffer (BCS) superconductivity.
Furthermore, unique opportunities exist for investigation of
interaction physics in the atomic system via techniques that have
been established for fundamental control over atomic interactions
\cite{control}.  In this Letter we present measurements of the
equilibrium thermodynamic properties and collisional dynamics of an
interacting Fermi gas composed of $^{40}$K atoms in two spin-states.

Producing a two-component Fermi gas, where both species are quantum
degenerate, builds on the earlier experimental procedure
\cite{sciarticle}.  Fermionic atoms are magnetically trapped in two
spin-states (magnetic sub-levels $m_f$=9/2 and $m_f$=7/2 in the
$^{40}$K $f=9/2$ hyperfine ground state) to allow the s-wave
collisions necessary for rethermalization during evaporative cooling
\cite{ourprl}. At low temperature, this mixture of spin-states is
stable against spin relaxation and $m_f$=9/2 and $m_f$=7/2 atoms can
only collide with each other \cite{ourpra}. Previously, the last
stage of evaporation slowly removed the $m_f$=7/2 atoms to
sympathetically cool the remaining $m_f$=9/2 atoms into the quantum
degenerate regime.  This last stage was eliminated by improving the
simultaneous evaporation of the two spin-state mixture so that an
equal mixture can now be cooled to $T/T_F\approx0.25$, where $T$ is
the temperature and $T_F$ the Fermi temperature. Improvements include
increased stability of the magnetic trapping field, elimination of
atoms in spin-states other than $m_f$=9/2 and $m_f$=7/2, and
increased microwave power for the evaporative knife. A technique has
also been implemented to simultaneously image both components in
order to extract information about the interacting gas. During the 10
ms expansion of the gas after release from the magnetic trap, a
Stern-Gerlach magnetic field with an 80 gauss/cm gradient is applied
to spatially separate atoms in the two spin-states \cite{ketterle}.
Each component is then resolved separately in absorption images
\cite{udnote} (see inset in Fig. 1).

The emergence of quantum degeneracy is observed through measurements
of the equilibrium thermodynamic properties \cite{sciarticle} of the
two-component gas.  In the quantum degenerate regime
($T/T_F\lesssim1$), the average energy per particle $E$ rises above
the classical expectation $3k_BT$. For the roughly equal ($46\%$
$m_f$=9/2) mixture of spin states used for evaporation, Fig. 1(a)
shows $E$ vs $T/T_F$ for each component. The excess energy
characteristic of quantum degenerate Fermi systems can clearly be
observed in both components.  For this data, $T$ and $E$ are
determined independently for each component from fits to absorption
images of the expanded gas. A fit to the Thomas-Fermi shape expected
for an ideal Fermi gas \cite{Rokshar} is used to measure $T$, while
$E$ is determined from a gaussian fit that is weighted to minimize
the fit deviation from the second moment of the image
\cite{energyfits}. The widths of both fits are adjusted by roughly
6$\%$ to account for distortions introduced by curvature in the
Stern-Gerlach field \cite{agnote}. The measured temperatures of the
two components match to within experimental uncertainty, as expected
for thermal equilibrium.

Thermodynamic data for a different spin mixture, $86\%$ $m_f$=9/2,
are displayed in Fig 1(b).  Here the spin composition is controlled
by removing some $m_f$=7/2 atoms after the bulk of the evaporation.
Changing the spin mixture manipulates the Fermi energies $E_F$ since
$E_F$ depends on the number of atoms $N$ through
$E_F=k_BT_F=\hbar\overline{\omega}(6N)^{1/3}$ \cite{Rokshar}, where
$\overline{\omega}=(\omega_r^2\omega_z)^{1/3}$ is the geometric mean
of the harmonic trap frequencies \cite{trapfreq}.  In the
thermodynamics for this less balanced spin mixture, again both
components reach quantum degeneracy.

For both mixtures, the thermodynamic data agree well with the ideal
gas theory prediction; this indicates that the mean-field energy due
to inter-particle interactions in the gas must be small compared to
the kinetic and potential energy. For Bose-Einstein condensates (BEC)
with similar number, temperature, and scattering lengths, the mean
field is quite significant. However, the influence of the mean field
for a Fermi gas is drastically reduced because the Fermi gas always
has higher energy and lower density than a Bose-Einstein condensate.
In fact, for our range of experimental conditions $E_{\rm
int}/k_BT_F<0.4\%$, where $E_{\rm int}$ is the interaction (mean
field) energy per particle \cite{intcalc}.

Also apparent in Fig. 1 is a misalignment of the corresponding
$m_f$=9/2 and $m_f$=7/2 points on the $T/T_F$ axis, indicating that
the two components are not equally degenerate. This is particularly
true for the 86$\%$ $m_f$=9/2 case where $E_F$ is roughly twice as
high for the $m_f$=9/2 component, and therefore the $m_f$=9/2
component is always more degenerate. Figure 2 displays the effect of
unequal $E_F$ by plotting the energy ratio $E_{9/2}/E_{7/2}$ vs
$T/T_{F, 9/2}$ ($T/T_F$ for the $m_f$=9/2 component). For the gas
with $86\%$ $m_f$=9/2 atoms, $E_{9/2}/E_{7/2}$ is measured as high as
1.4 in the quantum degenerate regime, strongly violating the
classical expectation $E_{9/2}/E_{7/2}=1$. However, when the gas has
roughly equal numbers of $m_f$=9/2 and $m_f$=7/2 atoms $E_F$ is
matched to within 13$\%$, and $E$ is roughly equal for both
components irrespective of $T/T_F$.

The observed imbalance in $E$ must arise from a change in the
collisional interactions in the gas since s-wave collisions would
normally redistribute energy equally between the two components.
Collisions are predicted to be suppressed by Pauli blocking, a
phenomenon common to all Fermi systems such as semiconductors, liquid
$^3$He, and nuclear matter.  At low $T/T_F$, the lowest energy states
of the trap are highly occupied and any collision resulting in a
final atom state at low energy is suppressed by the Pauli exclusion
principle.  The energy imbalance is then maintained because
collisions that remove energy from the more degenerate component are
the most strongly suppressed.

We have directly observed Pauli blocking of elastic collisions in
measurements of the thermal equilibration time.  The gas is taken out
of thermal equilibrium by a rapid removal of high energy atoms from
the $m_f$=7/2 component. Because of gravitational sag in the trap,
energy is preferentially removed from the radial direction. As a
result, the $m_f$=7/2 component is initially both out of
cross-dimensional equilibrium as well as out of equilibrium with the
$m_f$=9/2 component.  A sample data set showing the rethermalization
of the energy of the $m_f$=7/2 component is shown in Fig. 3. The
cross-dimensional relaxation rate can be used to obtain the elastic
collision cross-section \cite{ourprl,cross}. One can define an {\it
effective} collision cross-section $\sigma_{\rm eff}$ that
encapsulates the total effect of Pauli blocking on collisions
independent of changes in the density and temperature of the gas. The
value of $\sigma_{\rm eff}$ is predicted to vary from 0 at $T$=0 to
the s-wave cross-section $\sigma=4\pi a^2$ in the classical regime
\cite{galitskii}, where $a$ is the s-wave triplet scattering length.

Figure 4 shows the measured $\sigma_{\rm eff}$ vs $T/T_{F,\ 9/2}$ for
an 86$\%$ $m_f$=9/2 gas.  The rethermalization time constant $\tau$
is extracted from a fit to the time dependence of the $m_f$=7/2
component aspect ratio, assuming that the energy difference
$\delta=E_x-E_z$ relaxes exponentially ($E_z$ and $E_x$ refer to the
$m_f$=7/2 energy in the axial and one of the radial directions,
respectively). The cross-section is then determined through
$\frac{1}{\tau}=\frac{n\sigma_{eff}v}{\alpha}$, where
$n=\frac{1}{N_{7/2}}\int d^3\vec{r}\
n_{9/2}(\vec{r})n_{7/2}(\vec{r})$ is the overlap integral of the
density distributions, $v$ is the mean relative speed for a collision
between $m_f$=9/2 and $m_f$=7/2 atoms, and $\alpha$ is the average
number of collisions per $m_f$=7/2 atom required for
cross-dimensional equilibration. The product $nv$ is determined from
gaussian fits to the expanded images of each component
\cite{gaussian}.

For our data, the time dependence of the aspect ratio is complicated
by energy transfer from the $m_f$=9/2 component to the $m_f$=7/2
component. This is included in the fit to the aspect ratio by
assuming that the difference $\Delta=E_{9/2}-\eta E_{7/2}$ also
relaxes exponentially with time constant $\left(\frac{n\sigma_{\rm
eff}v}{A}\frac{N_{7/2}+\eta N_{9/2}}{N_{9/2}(1+\eta)}\right)^{-1}$,
where $\eta$ is the equilibrium ratio $E_{9/2}/E_{7/2}$, and $A$ is a
constant similar to $\alpha$. We have used a classical kinetic theory
calculation \cite{energyflow} to determine that $\alpha=A=0.75$.

In the classical regime, the measured value of $\sigma_{\rm eff}$
agrees, within the factor of two systematic uncertainty in atom
number, with the best known value of the scattering length for
$^{40}$K $a=169a_0$ \cite{s-wave}, where $a_0$ is the Bohr radius.
The effective cross-section drops by a factor of two at $T/T_{F,\
9/2}=0.2$. Within our uncertainty, the observed size of the Pauli
blocking effect agrees with the theoretical value of $\sigma_{\rm
eff}$ from a quantum kinetic calculation \cite{murraypaper} shown in
Fig. 4. In the calculation, the reduction in $\sigma_{\rm eff}$
represents the effect of Pauli blocking averaged over all possible
initial and final colliding atom states.

In conclusion, we have cooled an interacting Fermi gas of atoms to
one-fifth of the Fermi temperature.  At these low temperatures, we
have observed strong effects of the quantum statistics on the
thermodynamics and collisional dynamics of the gas. Pauli blocking is
predicted to cause other modifications in the behavior of the gas
including changes in the frequency and damping rate of shape
oscillations \cite{intcalc,Clark,shape} and the dipole
``scissors-mode" \cite{scissors} as well as suppression of light
scattering \cite{ourpra,light}.  In addition, future investigation
into the possibility of a paired state BCS-type phase in the atomic
Fermi gas will almost certainly require a two-component degenerate
system. Realizing this new phase will also require control over the
sign and strength of the interactions, which may be possible in
$^{40}$K via a predicted Feshbach resonance \cite{johnfesh}.

This work is supported by the National Institute of Standards and
Technology, the National Science Foundation, and the Office of
Naval Research.  The authors would like to express their
appreciation for useful discussions with C.E. Wieman, E.A.
Cornell, and M. Holland.

\begin{figure}  \epsfxsize=5 truein \epsfbox{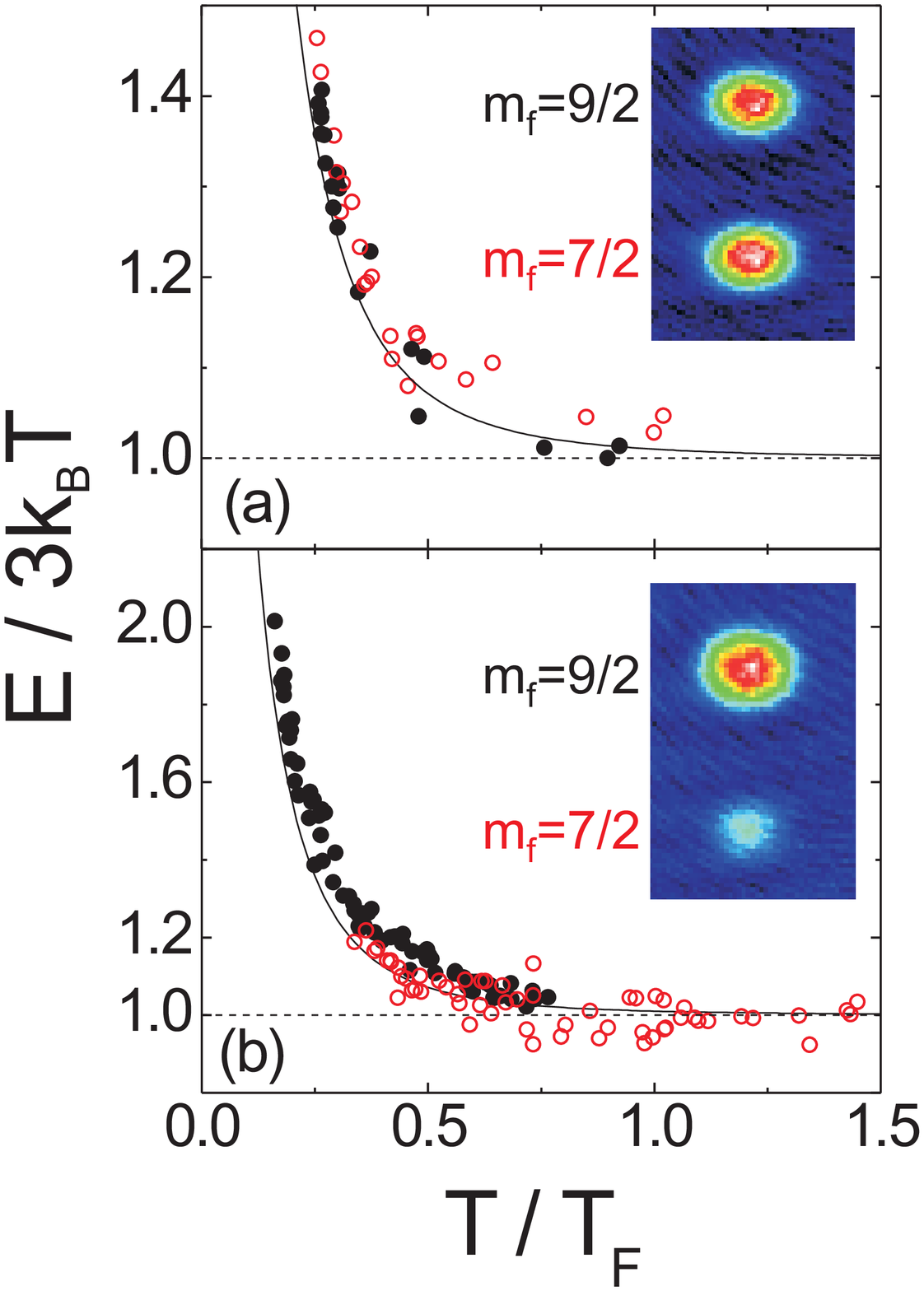}
\caption{Thermodynamics of the interacting gas.  The average energy
per particle $E$, extracted from absorption images such as the
examples shown in the insets, is displayed for two spin mixtures,
$46\%$ $m_f$=9/2 (a) and $86\%$ $m_f$=9/2 (b).  In the quantum
degenerate regime, the data deviate from the classical expectation
(dashed line) as the atoms form a Fermi sea arrangement in the energy
levels of the harmonic trapping potential. The data in (a) represent
the spin mixture used for evaporation, where we reach $T/T_F\sim0.25$
at 90 nK and $N=2.8\times10^5$ atoms. The data agree with the ideal
Fermi gas prediction for a harmonic trap, shown by the solid line.
Misalignment of corresponding $m_f$=9/2 and $m_f$=7/2 points on the
$T/T_F$ axis reflects a difference in the Fermi energies for the two
components. \label{fig1}}
\end{figure}

\begin{figure} \epsfxsize=5 truein \epsfbox{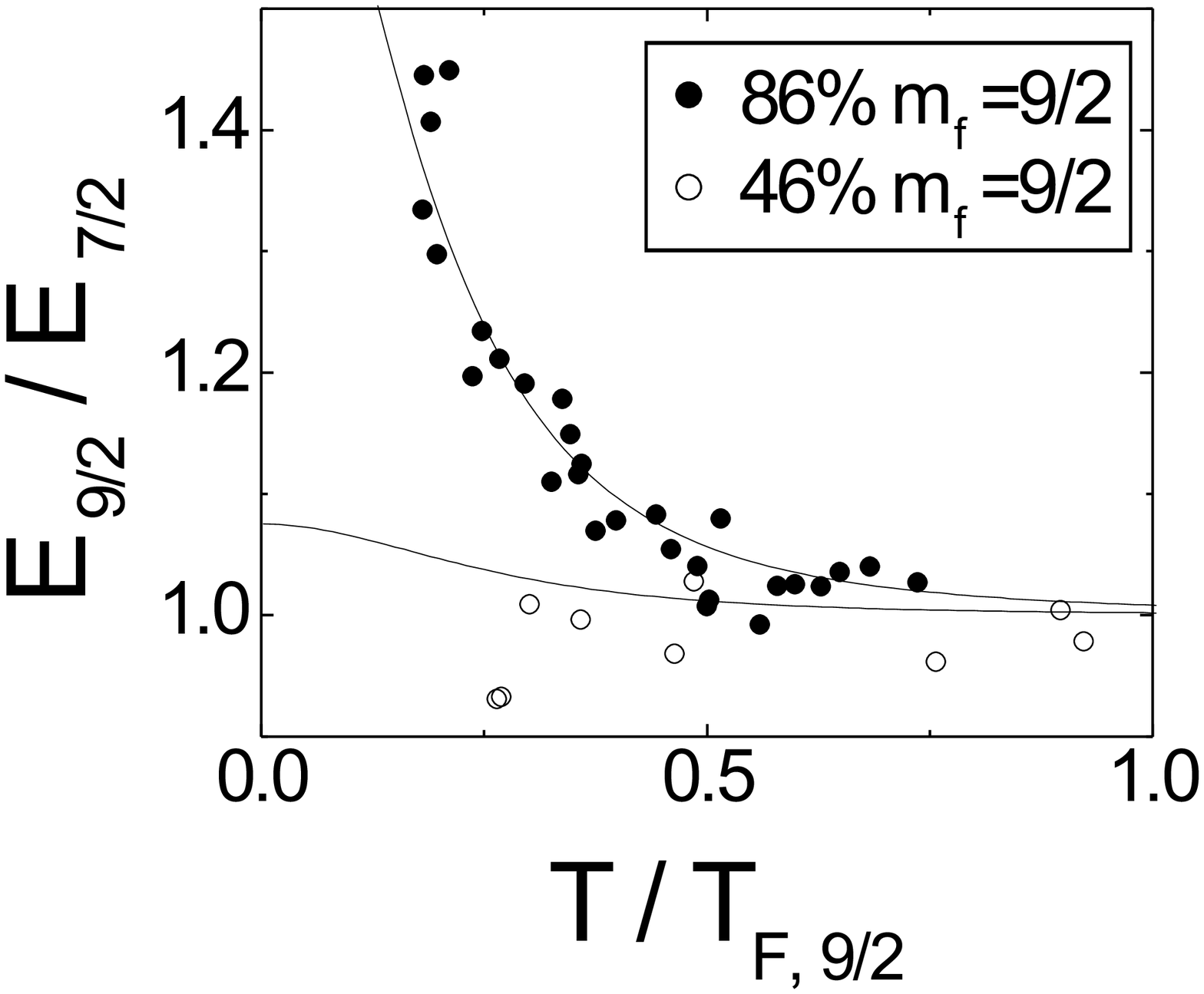}
\caption{Effect of Pauli blocking on the equilibrium thermodynamics
of the gas.  Using the same data shown in Fig. 1, the ratio of
energy, $E_{9/2}/E_{7/2}$, for pairs of clouds from each double image
is plotted vs. $T/T_{F,\ 9/2}$. Each point in this plot represents
the average of two runs of the experiment. For comparison, the
prediction for an ideal Fermi gas is shown by the solid lines.  The
energy imbalance revealed at low $T/T_F$ is maintained by Pauli
blocking of collisions. \label{fig2}}
\end{figure}

\begin{figure} \epsfxsize=5 truein \epsfbox{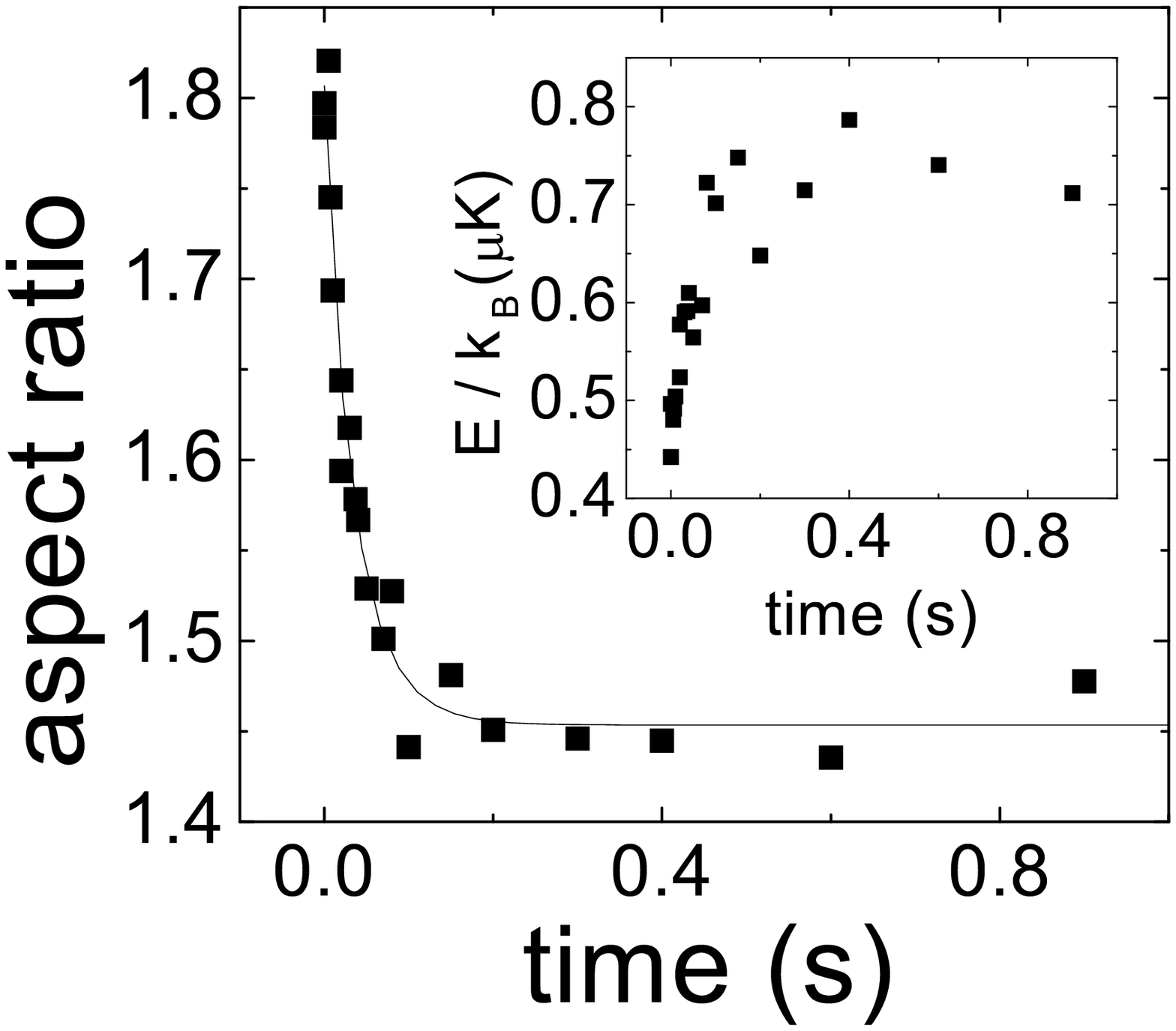}
\caption{Rethermalization data.  A plot of the aspect ratio (axial
size/radial size) of the expanded $m_f$=7/2 component vs. time shows
cross-dimensional energy rethermalization. The inset, a plot of the
average energy per particle of the $m_f$=7/2 component, shows the
simultaneous transfer of energy from the $m_f$=9/2 to $m_f$=7/2
component. The data in this figure were taken for a gas with
$T/T_{F,\ 9/2}$=0.5, $N_{9/2}=4.6\times10^5$, and
$N_{7/2}=7.7\times10^4$. A fit (solid lines) of the time dependent
aspect ratio is used to measure the effective collision
cross-section. \label{fig3}}
\end{figure}

\begin{figure} \epsfxsize=5 truein \epsfbox{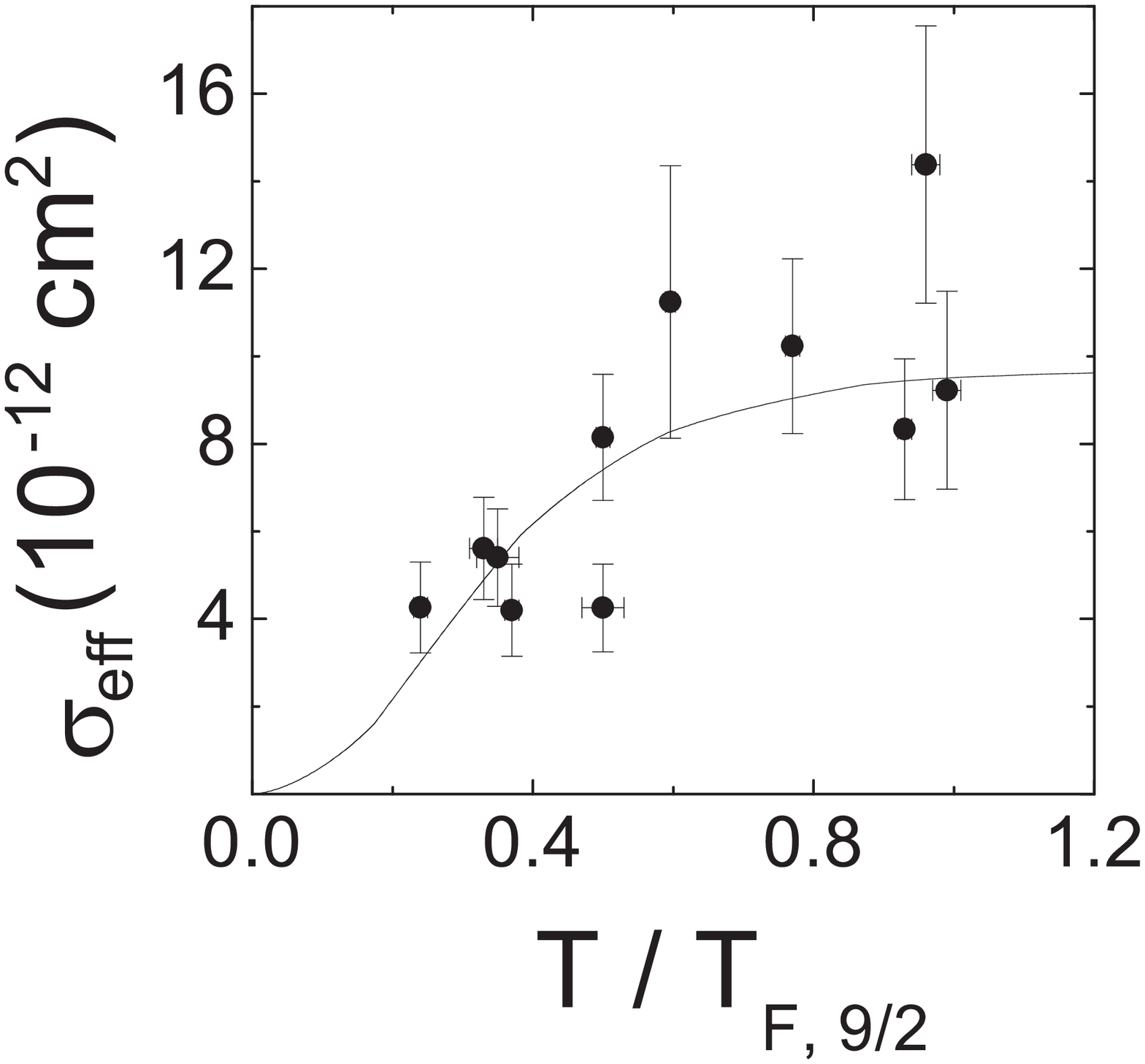}
\caption{Collisional Pauli blocking.  A factor of two decrease in the
effective elastic collision cross-section, $\sigma_{\rm eff}$, is
observed at low $T/T_F$. The error bars in $\sigma_{\rm eff}$ are
predominately from uncertainty in the fits to the time dependence of
the $m_f=7/2$ aspect ratio, while scatter in number and temperature
set the error bars in $T/T_{F,\ 9/2}$. In addition, there is a $26\%$
systematic uncertainty in $T/T_{F,\ 9/2}$ and a factor of two
systematic uncertainty in $\sigma_{\rm eff}$ from uncertainty in the
number determined from absorption imaging. The solid line shows the
result from a quantum kinetic calculation of the collision rate. At
high $T/T_F$ the data agree with the known value of the s-wave
collision cross-section, and at low $T/T_F$ the observed decrease in
$\sigma_{\rm eff}$ agrees with the quantum kinetic prediction.
\label{fig4}}
\end{figure}


\begin{references}
\bibitem{udnote} A controlled mixture of laser polarization
is used for resonant absorption imaging in order to avoid lensing
caused by the high density gradients present in the gas.
\bibitem{energyfits}  We have checked that the weighted
fit returns the correct value of $E$ for calculated Fermi-Dirac,
Bose-Einstein, and classical distributions.  We measure the expansion
energy per particle, and plot $E$ (twice the expansion energy)
assuming that mean field effects are small \cite{Clark}.
\bibitem{agnote} The correction factor is measured across the range
of temperatures used in this experiment and is consistent with a
classical calculation describing the individual trajectories of
atoms experienced during the expansion.
\bibitem{trapfreq} The two spin-states have different magnetic moments
and therefore experience slightly different harmonic oscillator
frequencies.  The magnetic trap frequencies for all of the
experiments described in this Letter are 134.0 Hz radial and 19.9 Hz
axial for the $m_f$=9/2 atoms.  The corresponding frequencies for the
$m_f$=7/2 atoms are reduced by the square root of the ratio of the
magnetic moments ($\sqrt{7/9}$).
\end{references}
\end{document}